\documentclass[aps,pre,showpacs,twocolumn,groupedaddress,superscriptaddress]{revtex4}
\usepackage{graphicx}%
\usepackage{amsmath}
\usepackage{dcolumn}

\begin{document}

\title[Event-related (De)synchronization]{A general coupled nonlinear oscillator model for event-related (de)synchronization}

\author{Jane H. Sheeba, V. K. Chandrasekar and M. Lakshmanan}
\affiliation{Centre for Nonlinear Dynamics, School of Physics,
Bharathidasan University, Tiruchirapalli - 620 024, India}

\begin{abstract}
Changes in the level of synchronization and desynchronization in coupled oscillator systems due to an external stimulus is called event related synchronization or desynchronization (ERS/ERD). Such changes occur in real life systems where the collective activity of the entities of a coupled system is affected by some external influence. In order to understand the role played by the external influence in the occurrence of ERD and ERS, we study a system of coupled nonlinear oscillators in the presence of an external stimulus signal. We find that the phenomena of ERS and ERD are generic and occur in all types of coupled oscillator systems. We also find that the same external stimulus signal can cause ERS and ERD depending upon the strength of the signal. We identify the stability of the ERS and ERD states and also find analytical and numerical boundaries between the different synchronization regimes involved in the occurrence of ERD and ERS.
\end{abstract}


\maketitle

\section{Introduction}
Synchronization is an ubiquitous natural phenomenon that occurs widely in real systems including those in physics \cite{Smet:09}, chemistry \cite{Kuramoto:84,Kiss:01, Kiss:02}, biology \cite{Winfree:84,Mosekilde:02, Neiman:02} and nano-technology \cite{Kaka:05,Slavin:09}. This phenomenon is an active topic of current research and is being extensively studied \cite{Strogatz:00,Pikovsky:01,Barahona:02,Rosenblum:03,Sorrentino:08}. Nevertheless, synchronization is not always a desirable phenomenon. In some cases it is desirable while in some other times it is undesirable and hence a mechanism to desynchronize becomes necessary for normal behavior. For example, synchronization is desirable in the cases of lasers and Josephson junction arrays \cite{Cawthorne:99,Tsygankov:02}, coupled spin torque nano-oscillators where coherent microwave power is needed \cite{Mohanty:05, Grollier:06}, and in the brain when synchronization of neuronal oscillations  facilitate cognition via temporal coding of information \cite{Fries:05,Yamaguchi:07,Sheeba:08a}. On the other hand, synchronization is undesirable when pedestrians walk on the Millennium Bridge 
\cite{Strogatz:05} and when mass synchronization of neuronal oscillators occurs at a particular frequency band resulting in pathologies like trauma, Parkinson's tremor and so on \cite{Hammond:07}. Apart from these, synchronization or desynchronization of neuronal oscillations in the brain are found to facilitate task selection and performance. When such a synchronization or desynchronization occurs in the brain or other physical systems mentioned above due to a particular task or an event or an influence of an event then it is called an event-related synchronization (ERS) or desynchronization (ERD). 

In this direction in order to understand the phenomena of ERD and ERS, we study the synchronization and desynchronization dynamics of a system of coupled nonlinear oscillators due to the effect of the strength of the external stimulus which represents some external influence. In particular, we demonstrate the occurrence of ERD and ERS due to the change in the strength of the external stimulus. We also find the occurrence of ERD/ERS in periodic and chaotic systems and also with different forms of coupling. We therefore find that the occurrence of ERD and ERS due to external influence is generic and is not limited to the type of the system or the coupling. 

The plan of the article is as follows. In the following section we present a general model of coupled nonlinear oscillators in the presence of an external influence to explain ERD/ERS. We discuss the case of the limit cycle Stuart-Landau (SL) oscillators in Sec. \ref{SL} as an illustration of our model. We quantify the strength of ERD/ERS by defining the intensity $I$ and identify the analytical stability boundaries of the ERD and ERS states. Further we find out the stability of the ERD and ERS states and also identify the analytical and numerical bifurcation boundaries between the different synchronization regimes. We also demonstrate the occurrence of ERD and ERS in the presence of other forms of couplings like weighted coupling and time delayed coupling. In Sec. \ref{chaotic} we show that ERD/ERS phenomena also occur in chaotic systems by illustrating the case of coupled R\"{o}ssler oscillators. We point out potential applications of the model and the results in Sec. \ref{App}. Finally we present our conclusions in Sec. \ref{conc}.

\section{The general model for ERD/ERS}
\label{model}
Let us consider a system of coupled limit cycle nonlinear oscillators \cite{Kuramoto:84,Tass:95,Pikovsky:01,Rosenblum:04,Tukhlina:07,Kiss:08,Sheeba:08a} subject to \emph{external stimulus}. The external stimulus represents a dynamic signal from outside the system (brain or or other physical systems such as spin torque nano oscillators) or from a distant region inside the system and typically represents the influence of the event (and not necessarily the actual event itself). With this in consideration, the model equations shall be cast into the following form,
\begin{eqnarray}
\dot{{\bf X}_j}&=& {\bf F}({\bf X}_j,{\bf \epsilon}_j) + \frac{A}{N}
\sum_{k=1}^{N} ({\bf X}_k-{\bf X}_j) + B {\bf Y},\quad (\; \dot{}=\frac{d}{dt})\nonumber \\
\dot{{\bf Y}}&=&{\bf G}({\bf Y},{\bf \epsilon}_e),
\label{SL01}
\end{eqnarray}
where $j=1,2, \ldots, N$. Here ${\bf F}({\bf X}_j,{\bf \epsilon}_j)$ represents the nonlinear limit cycle behavior of the $j$th uncoupled oscillator and ${\bf \epsilon}_j$ is the corresponding system parameter. $A$ is the coupling strength between the oscillators in the system and $B$ represents the coupling strength between the oscillators in the system and the external stimulus. Our findings on the occurrence of ERD in this model has been briefly reported in \cite{Sheeba:09}.

From a detailed analysis of generic models of the form (\ref{SL01}), we have found that the external stimulus $B$ typically affects synchronization in the system of coupled oscillators. If the system of coupled oscillators is completely synchronized, that is, ${\bf X}_j={\bf X}$, $j=1, 2, \ldots , N$ (in this state all the oscillators behave as one) due to the strength of the coupling $A$ (when $B=0$), then a sufficient strength of $B$ causes desynchronization (ERD, since external stimulus causes desynchronization) in the system. Typically ERD occurs due to a small group of oscillators that separate themselves from the larger synchronized group. The separated small group either remains desynchronized or synchronizes itself to a frequency different from that of the larger synchronized group. Both of these cases will be demonstrated in the following Sections. 

The above behavior is generic and can also occur in chaotic systems leading to chaotic synchronization/desynchronization. Further, this behavior occurs irrespective of the form of coupling (weighted, nonlocal, etc.) and also in delay coupled systems. In all the cases, when the system is synchronized due to the strength of the coupling and when the strength of the external stimulus is sufficient, ERD/ERS occurs.  In the following we present our theoretical findings and discuss the qualitative connection between our findings and experimental observations.

\section{ERD/ERS in a system of Stuart-Landau oscillators}
\label{SL}
A well-known and well-studied limit cycle nonlinear oscillator is the Stuart-Landau (SL) oscillator \cite{Kuramoto:84}. A system of coupled SL oscillators is described by the following set of coupled complex first order nonlinear ordinary differential equations (ODEs),
\begin{eqnarray}
\dot{z}_j&=& (a+i\omega_j-(1+ic)|z_j|^2)z_j+ \frac{A}{N}\sum_{k=1}^{N} (z_k-z_j) + B z_{e},\nonumber\\
\dot{z}_e&=& (a_e+i\omega_e-(1+ic_e)|z_e|^2)z_e.
\label{SL02}
\end{eqnarray}
Here $c$ is the nonisochronicity parameter, $a$ is the Hopf bifurcation parameter and $\Gamma_{k,j} =(z_k-z_j)$ is the coupling function. $z_j=x_j+iy_j$ is the complex amplitude of the $j$th oscillator with natural frequency $\omega_j$. $z_{e}$ is another SL oscillator with natural frequency $\omega_e$ and nonisochronicity parameter $c_{e}$.
Without coupling $(A,B=0)$, the dynamical equation of an individual Stuart-Landau oscillator becomes
\begin{eqnarray}
\dot{r}= (a-r^2)r, \quad \dot{\theta}=\omega-cr^2,
\label{SL02a}
\end{eqnarray}
where $z=re^{i\theta}$. The fixed points of the radial equation are $r_0=0$ and $\sqrt {a}$. From a linear stability analysis the eigenvalue is found as $\lambda=(a-3r_0^2)$. For $a<0$ the system has a stable fixed point $z=0$ which is a linearly stable state corresponding to amplitude death. For $a>0$, the fixed point $z=0$ becomes linearly unstable and a limit cycle oscillation is established with radius $r=\sqrt {a}$ and phase $\theta=(\omega-ac)t+\theta_0$, where $\theta_0$ is a constant. For more detailed studies about synchronization in populations of SL oscillators one may refer to \cite{Popovych:06,Teramae:04}.

\begin{figure}
\begin{center}
\includegraphics[width=6.0cm]{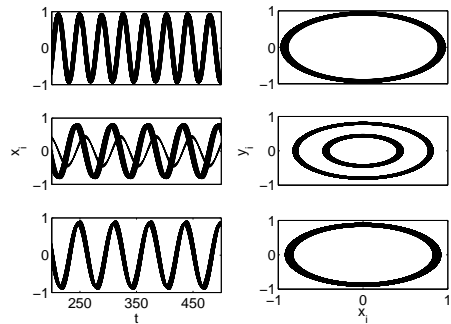}
\caption{Occurrence of ERD and ERS in a system of 1000 coupled Stuart-Landau oscillators for $A=1.1$, $c=1.5$, $c_{e}=2.0$, $a=a_e=1.0$, $\omega_e=1.5$,$\gamma=0.005$ and $\bar{\omega}=0.5$ in equation (\ref{SL02}). Left column shows the time evolution of the real part of the state vectors, $x_i$, and the right column depicts the corresponding phase portraits on the ($x_i$,$y_i$) plane. Here (top) $B=0.01$, (middle) $B=0.08$ and (bottom) $B=0.35$. Note $z_i=x_i+y_i$.}

\label{SL1}
\end{center}
\end{figure}

\begin{figure}
\begin{center}
\includegraphics[width=6.0cm]{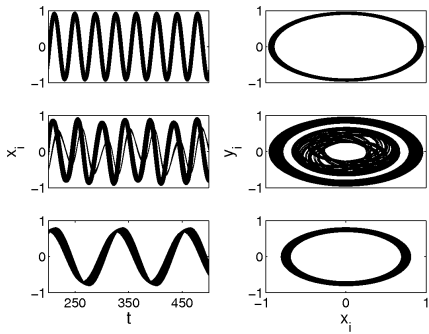}
\caption{The case of quasi-periodic synchronization in the ERD state in a system  of 1000 coupled Stuart-Landau oscillators. Here $\omega_e=1.0$; all the other parameter values are the same as in Figure \ref{SL1}.}
\label{SL2}
\end{center}
\end{figure}

The occurrence of ERD and ERS in a system of 1000 coupled Stuart-Landau oscillators is numerically demonstrated in Figure \ref{SL1} for a Lorentzian distribution of natural frequencies $g(\omega)=\frac{\gamma}{\pi}[\gamma^2+(\omega-\bar{\omega})^2]^{-1}$, where $\gamma$ is the half width at half maximum and $\bar{\omega}$ is the central frequency. For the given set of system parameters ($A=1.1$, $c=1.5$, $c_{e}=2$, $\omega_e=1.5$, $\gamma=0.005$, $\bar{\omega}=0.5$), ERD occurs when moving from the top to the mid panel and ERS occurs when moving from the mid to the bottom panel, upon increasing $B$. The time evolution of the real part of the state vectors, $x_i$, and the corresponding phase portraits are plotted in the left and the right columns, respectively. When $B=0.01$ all the oscillators are synchronized in phase (while there is slight desynchronization in the amplitude) in the top panel. On increasing $B$ from $0.01$ to $0.08$, in the mid-panel, ERD occurs when some oscillators separate themselves from the synchronized group and are synchronized to a different frequency. Even though the separated group of oscillators are synchronized among themselves, we refer to this state as being desynchronized because from the point of view of the entire system, this is desynchronization. Further due to this separation, the intensity or the strength of synchronization (discussed in detail in the following subsection) of the original synchronized group (shown in the top panel) is reduced. Now, in this state, when we increase $B$ from $0.08$ to $0.35$ ERS occurs (with reference to the mid-panel) when the separated group of oscillators again synchronize with the other oscillators. The separated group of oscillators need not always be synchronized but can be either desynchronized or quasi-periodically synchronized. The case of quasi-periodic synchronization is shown in Figure \ref{SL2}. Here ERD occurs in the mid-panel for the same values of (system and control) parameters as in Figure \ref{SL1}, except that now $\omega_e=1.0$, where the separated group is quasi-periodically synchronized. 

\subsection{Measure of ERD/ERS}
The strength or intensity of synchronization can be measured by the number of oscillators that are oscillating in synchrony. The more the number of oscillators that are oscillating in synchrony the more will be the strength of synchronization. Hence an increase in the strength or intensity of synchronization denotes the occurrence of ERS and a decrease in the intensity denotes the occurrence of ERD. With this reasoning we quantify the intensity of synchronization using the phases of the oscillators by defining the quantity \cite{Kuramoto:84,Acebron:05}
\begin{eqnarray}
I=<\bar{|e^{i\theta_j}|}>=\frac{1}{T}\int_{0}^{T}\bigg[\mid \frac{1}{N}\sum_{j=1}^{N}e^{i\theta_j}\mid\bigg]dt,
\end{eqnarray}
where $\theta_j=\tan^{-1}(y_j/x_j)$ is the phase of the $j$th oscillator. Here the bar represents average over all oscillators in the population and the angle brackets represent time average. In the state of complete desynchronization $I=0$ and for complete synchronization $I=1$ (while we neglect very little desynchronization in the amplitude). For partial synchronization $I$ takes a value between 0 and 1; the more oscillators are oscillating in synchrony the higher will be the value of $I$. We use \emph{phase} desynchronization/synchronization to characterize the occurrence of ERD/ERS because, upon increasing $B$ while the oscillators are in complete synchronization, very little amplitude desynchronization occurs first and then for an increase in $B$ desynchronization occurs in the phase and the desynchronization in the amplitude is also increased (ERD) (see for example Figures \ref{SL1} and \ref{SL2}). Further increase in $B$ brings back synchronization in the phase with slight desynchronization in the amplitude (ERS). Therefore monitoring the intensity of phase synchronization will facilitate monitoring the occurrence of ERD/ERS.

\begin{figure}
\begin{center}
\includegraphics[width=6.0cm]{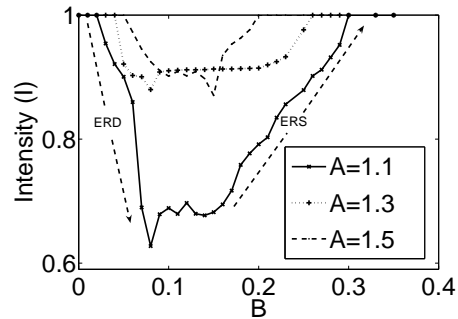}
\caption{Change in the intensity $I$ for the SL system (\ref{SL02}) for varying stimulus strength $B$ for different values of the coupling strength $A$. All other parameters are the same as in Figure \ref{SL2}. In the direction of increasing $B$, downward arrow (that is, a decrease in $I$) denotes ERD and the upward arrow (that is, an increase in $I$) denotes the occurrence of ERS.}
\label{SLint1}
\end{center}
\end{figure}

Figure \ref{SLint1} shows that there exists a critical strength of the external stimulus for ERD/ERS to occur. Upon varying the stimulus strength $B$, ERD occurs when there is a decrease in the intensity at a sufficient $B$ (as denoted by a downward arrow in Figure \ref{SLint1}) and when there is an increase in the intensity for sufficient $B$, ERS occurs (as denoted by an upward arrow in Figure \ref{SLint1}). It may also be noted that the ERD/ERS occurs for a finite window of the stimulus strength which depends upon the coupling strength $A$. When we start with a state where the system is completely phase synchronized due to the coupling strength $A$, the ERD/ERS occurrence window decreases with stronger $A$. This means that for very strong coupling $A$ there is a very less chance that ERD/ERS occurs.

\subsection{Stability of the ERD/ERS state}

Our numerical simulations show that when ERD occurs for a sufficient strength of the stimulus, the synchronized group splits into one major synchronized group and a small separated group (synchronized or desynchronized) comprising of few oscillators (say $N_2$) compared to the size of the major synchronized group (say $N_1$). Therefore the size ratio of this two cluster state is $r:1-r$ where $r=N_1/N$ and $N=N_1+N_2$, and in the desynchronized state $r>>0$ ($\sim 1$). For the case of SL oscillators (\ref{SL02}) with an external force $z_{e}=e^{i\omega_e t}$ (so that $c_e=0$ and $a_e=1$), the dynamics of the large group $z_j=u,\;\;j=1,2,\ldots N_1$ and the small group $z_j=v,\;\;j=N_1+1,N_1+2,\ldots N$  can be written as
\begin{eqnarray}
\dot{u} &\approx &((a+i\omega)-(1+ic)|u|^2)u+B e^{i\omega_e t},\nonumber\\
\dot{v} &\approx & ((a-A+i\omega)-(1+ic)|v|^2)v+Au+Be^{i\omega_e t}.
\label{anal01}
\end{eqnarray}
If we assume that the major synchronized group is completely synchronized with the external force $z_{e}=e^{i\omega_e t}$, while the small group is not, we get $\omega=\omega_e+c$ and $a=1-B$. From this reasoning, the dynamics of the small group can be written in the form
\begin{eqnarray}
\dot{w}= ((a-A+ic)-(1+ic)|w|^2)w+A+B,
\label{anal02}
\end{eqnarray}
where $w=ve^{-i\omega_e t}$. The occurrence of ERD/ERS can be understood from the fixed points of this equation \cite{Sheeba:09}. Due to the cubic nature of equation (\ref{anal02}) it has three fixed points which determine the stability of the ERD/ERS state. One of the fixed points, namely $w_1=1$, is stable. The eigenvalues of the linearized version of (\ref{anal02}) are $\lambda_{1,2}=-(A+B), -(2+A+B)$. This corresponds to a complete synchronization (synchronization of all the oscillators) of the population, that is $u=v$ \cite{Daido:07}. The other two fixed points (say $w_2$ and $w_3$) determine the stability of the desynchronized state. The fixed points $w_2$ and $w_3$ exist for
\begin{eqnarray}
B<B_I=(1+c^2)(\sqrt{1+c^2}-1)/(2c^2)-A,
\label{anal03}
\end{eqnarray}
where $B_I$ is a saddle-node bifurcation point and the fixed points are given by
\begin{eqnarray}
w_{2,3}&=&\frac{1}{2(1+c^2)}\bigg(1+c^2-2(A+B)
\nonumber\\&&\pm\sqrt{(1+c^2-2(A+B))^2-4(1+c^2)(A+B)^2}\bigg).
\label{anal06}
\end{eqnarray}
Performing a linear stability analysis on equation (\ref{anal02}) we find that the determinant of the Jacobian matrix for $w_2$ and $w_3$ is
\begin{eqnarray}
\mbox{det} (J_{w_{2,3}}) =\frac{\mp 2(A+B)(A+B+2)\triangle}{(1+c^2+2(A+B))\pm \triangle},
\label{anal07}
\end{eqnarray}
where $\triangle=((1+c^2+2(A+B))^2-4(1+c^2)(A+B)(A+B+2))^{\frac{1}{2}}$. It is easily checked that the determinant (\ref{anal07}) is always negative (positive) for $w_2$ ($w_3$). Therefore the fixed point $w_2$ turns out to be a saddle. The other fixed point $w_3$ is either an unstable node or focus for $|c|\leq1$. When $|c|>1$ and $B>B_{II}$, the fixed point $w_3$ is either a stable node or focus; here $B_{II}$ is a Hopf bifurcation point given by
\begin{eqnarray}
B_{II}=(1+c^2)/(\sqrt{4+(1+c^2)^2}+2)-A,
\label{anal04}
\end{eqnarray}
which is determined from the condition $\mbox{tr}(J_{w_3})=0$. Here
\begin{eqnarray}
\mbox{tr}(J_{w_3}) =2\frac{(1-c^2)(A+B)+\triangle}{(1+c^2)}.
\label{anal07a}
\end{eqnarray}

\begin{figure}
\begin{center}
\includegraphics[width=8.0cm]{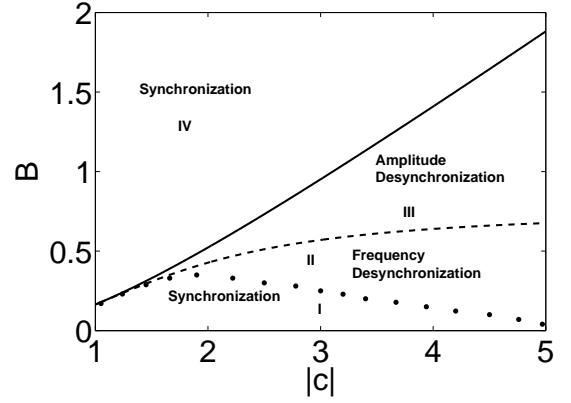}
\caption{$B-|c|$ phase diagram. $B_I$ (solid line) and $B_{II}$ (dashed line) are the analytically obtained saddle-node and Hopf bifurcation boundaries respectively. The boundary $B_{III}$ (dotted line) is obtained numerically by solving the evolution equation for $w$ given in the text.}
\label{SLanal1}
\end{center}
\end{figure}

Further for $B<B_{III}$ only synchronized solutions are stable and desynchronized solutions do not exist. For given values of parameters the bifurcation boundaries are plotted in Figure \ref{SLanal1}. In the regions I and IV all the oscillators are synchronized, while in the region II there is frequency desynchronization (mid-panel of Figure \ref{SL2} is an example since the small group oscillates quasi-periodically while the major group oscillates periodically) between the two clusters and in the region III there is amplitude desynchronization between the two clusters. At the point $B_{III}$, both the saddle and the Hopf bifurcation points merge and disappear and this point is called a saddle-connection point. It is obtained numerically by solving equation (\ref{anal02}) and is shown as a dotted line in Figure \ref{SLanal1}. Thus desynchronized solutions exist in the region between $B_I$ and $B_{III}$. This is in agreement with our numerical observations. A movement in the parameter space in the direction $I\rightarrow II$ or $IV\rightarrow III$ represents ERD and a movement in the opposite direction represents ERS.

\begin{figure}
\begin{center}
\includegraphics[width=7.50cm]{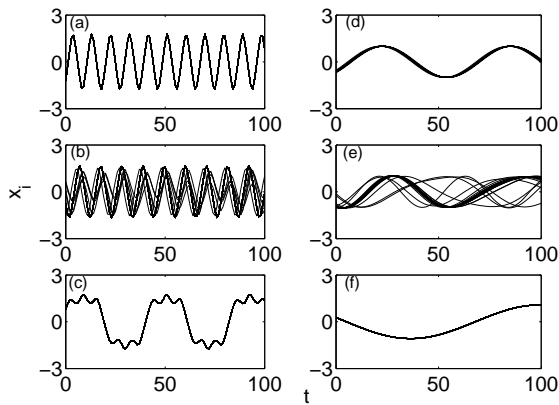}
\caption{The time evolution of the state vectors $x_i$ of a system of 100 delay coupled SL oscillators showing the occurrence of ERD/ERS in panels (a) - (c). Here $\tau=0.1$, $A=0.035$, $c=2.5$, $c_{e}=3.0$, $a=a_e=1.0$, $\omega_e=1.5$,$\gamma=0.0005$ and $\bar{\omega}=1.5$ and $B=0.0$ for (a), $B=0.05$ for (b) and $B=0.4$ for (c). Occurrence of ERD/ERS in a system of 100 coupled SL oscillators with random weighted coupling with weight factor 3.5 is shown in panels (d)-(f). Here $c=2.5$, $c_{e}=1.0$, $\omega_e=0.5$,$\gamma=0.005$ and $\bar{\omega}=1.5$ and $B=0.0$ for (d), $B=0.1$ for (e) and $B=1.1$ for (f).}
\label{SL_delay}
\end{center}
\end{figure}

\subsection{ERD/ERS in systems with other forms of coupling}
\label{other}
In order to demonstrate further the validity of the model, we first consider a system of delay coupled SL oscillators described by
\begin{eqnarray}
\dot{z}_j&=& (a+i\omega_j-(1+ic)|z_j|^2)z_j
\nonumber\\&&\qquad+ \frac{A}{N}\sum_{k=1}^{N} (z_k(t-\tau)-z_j) + B z_{e},
\label{SL02de}
\end{eqnarray}
where $\tau$ is the time delay parameter. We simulate (\ref{SL02de}) with $\tau=0.1$ and find the occurrence of ERD/ERS as depicted in Figure \ref{SL_delay} (a)-(c) for  $A=0.035$, $c=2.5$, $c_{e}=3.0$, $\omega_e=1.5$,$\gamma=0.0005$ and $\bar{\omega}=1.5$. In panel (a) the oscillators in the system are synchronized due to the strength of coupling $A$ when the external stimulus is absent. In panel (b) when the external stimulus is switched on with a strength $B=0.05$ ERD occurs in the system where a few of the oscillators separate themselves from the major synchronized group. When the strength of the external stimulus is further increased to $B=0.4$ in panel (c) ERS occurs in the system and all the separated oscillators get back to synchronization with the major group. Secondly, ERD/ERS also occurs in a system with random weighted coupling described by
\begin{eqnarray}
\dot{z}_j&=& (a+i\omega_j-(1+ic)|z_j|^2)z_j
\nonumber\\&&\qquad+ \sum_{k=1}^{N}A_{jk}(z_k-z_j) + B z_{e},
\label{SL02we}
\end{eqnarray}
where $A_{ij}$ is the random coupling matrix. Figure \ref{SL_delay} (d)-(f) demonstrates the occurrence of ERD (panel (e)) and ERS (panel (f)) in this system due to the stimulus strength. Thus we find that the form of the coupling is immaterial in causing the occurrence of ERD/ERS. For the non-delayed case, if we reduce our model to a phase model the bifurcations match with those presented in \cite{Sakaguchi:88,Childs:08,Ott:08} where the authors have studied a system of coupled phase oscillators in the presence of external forcing.

\section{ERD/ERS in a system of coupled chaotic oscillators}
\label{chaotic}
The occurrence of ERD/ERS is generic and not limited to limit cycle oscillator systems alone. For example let us consider a system of coupled R\"{o}ssler oscillators in the presence of an external field ($B\neq 0$)
\begin{eqnarray}
\dot{x}_j&=&-\omega_jy_j-z_j + B x_e,
\nonumber\\\dot{y}_j&=&\omega_jx_j+ay_j+ \frac{A}{N}
\sum_{k=1}^{N} (y_k-y_j),\nonumber\\
\dot{z}_j&=&b+z_j(x_j-c),
\label{Ro01}
\end{eqnarray}
where the external stimulus is described by
\begin{eqnarray}
\dot{x}_e=-\omega_ey_e-z_e,\; \dot{y}_e=\omega_jx_e+ay_e,\;\dot{z}_e=b_e+z_e(x_e-c_e).\nonumber
\label{Ro01a}
\end{eqnarray}
Also this system is found to exhibit ERD/ERS. Here $a$, $b$, and $c$ are the system parameters that determine the periodicity/chaoticity of the system. A Lorentzian distribution of the natural frequencies $\omega_j$ are chosen for numerical simulation. The external stimulus is represented by $x_e$ and in this case we have introduced coupling with the $x$ component of the external stimulus. However, the phenomenon demonstrated here can also be observed irrespective of the component in which the external stimulus is coupled to. The system parameters of the coupled R\"{o}ssler oscillators are chosen so that they are in the periodic regimes while those of the external oscillator are chosen so that it is in the chaotic regime.

\begin{figure}
\begin{center}
\includegraphics[width=7.50cm]{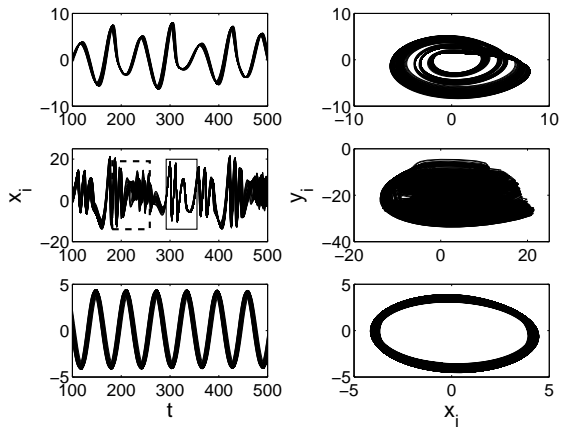}
\caption{Occurrence of ERD and ERS in a system of 1000 coupled R\"{o}ssler oscillators. Here $A=0.7$, $a=0.1$, $b=0.5$, $c=4.0$ and $a_e=0.1$, $b_e=0.1$, $c_e=9.0$, $\gamma=0.005$, $\bar{\omega}=1.0$ and (top) $B=0.1$, (mid) $B=1.8$ and (bottom) $B=2.3$. Left column shows the time evolution of the state vectors $x_i$ and the right column depicts the corresponding phase portraits on the ($x_i$,$y_i$) plane. A blow up of the dotted and solid rectangular regions marked in the mid-panel is shown in Figure \ref{Ross_zoom}.}
\label{Ross_ERD1}
\end{center}
\end{figure}

\begin{figure}
\begin{center}
\includegraphics[width=6.50cm]{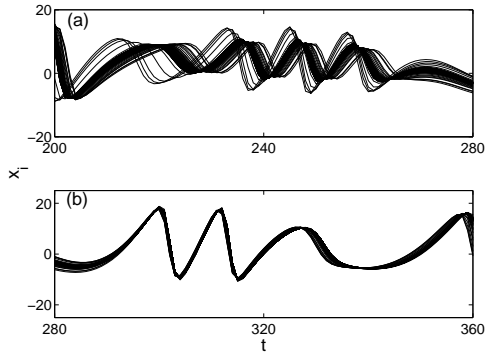}
\caption{A blow up of the dotted and the solid rectangular regions marked in Figure \ref{Ross_ERD1} is shown in panels (a) and (b), respectively. In panel (a) the separated group is desynchronized while in (b) it is synchronized with the major group.}
\label{Ross_zoom}
\end{center}
\end{figure}

An illustration of ERD and ERS in this system of 1000 coupled oscillators is depicted in Figure \ref{Ross_ERD1} for a Lorentzian distribution of natural frequencies with $\bar{\omega}=1$, $a=0.1$, $b=0.5$, $c=4.0$ and $a_e=0.1$, $b_e=0.1$, $c_e=9.0$ and $\gamma=0.005$. The top panel is the state of complete phase synchronization due to the coupling strength $A=0.7$ when $B=0.1$. ERD occurs in the mid-panel for $B=1.8$. In this state some of the oscillators separate from the synchronized group and are desynchronized. However unlike the case of ERD in SL oscillators (depicted in Figure \ref{SL1}) this ERD state is not stable but breathes in time. That is the separated group is desynchronized at certain times while it is synchronized with the major group at other times. This is obvious from Figure \ref{Ross_zoom} which is a zoom of the rectangular regions marked in Figure \ref{Ross_ERD1}. Figure \ref{Ross_zoom} (a) corresponds to the dotted rectangular region, which is the state where some oscillators are separated and desynchronized from the major synchronized group. Figure \ref{Ross_zoom} (b) corresponds to the solid rectangular region, which is the state where all the oscillators are synchronized. We have confirmed that this is not a transient since we have plotted the data after allowing a sufficiently large number of iterations and the breathing phenomenon is found to occur periodically. Thus ERD in this state is not stable but time varying. On further increase in the strength of the stimulus to $B=2.3$, ERS occurs in the system when all the oscillators are synchronized.

\begin{figure}
\begin{center}
\includegraphics[width=7.50cm]{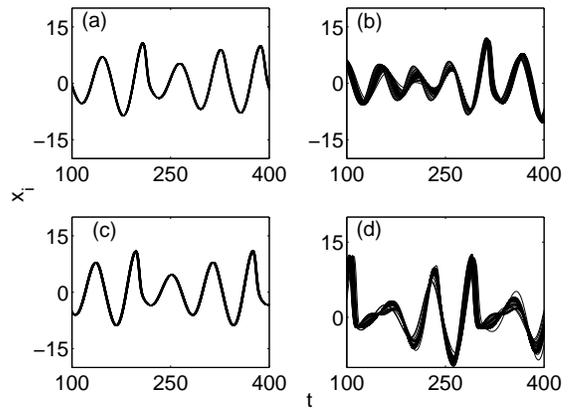}
\caption{Occurrence of ERD in a system of 1000 coupled R\"{o}ssler oscillators that are in the chaotic regime. In panels (a) and (b) the external oscillator is also chaotic with $a_e=0.2$, $b_e=0.2$, $c_e=5.7$. In panels (c) and (d) the external oscillator is periodic with $a_e=0.1$, $b_e=0.5$, $c_e=4.0$. ERD occurs in panels (b) and (d). Here $A=0.7$, $a=0.2$, $b=0.2$, $c=5.7$, $\gamma=0.005$, $\bar{\omega}=1.0$ and $B=0.01$ for (a), $B=0.15$ for (b), $B=0.03$ for (c) and $B=0.21$ for (d).}
\label{Ross_new}
\end{center}
\end{figure}

It may be noted that for illustrative purpose we have shown here the occurrence of ERD/ERS in a system of coupled R\"{o}ssler oscillators that are in periodic regime stimulated by a chaotic R\"{o}ssler oscillator. However, the same can be observed even if the external oscillator is in the periodic regime and coupled oscillators are in the chaotic regime and also for cases where the coupled oscillators and the external oscillator are both in chaotic or periodic regimes. For illustration we have shown the occurrence of ERD in a system of coupled chaotic R\"{o}ssler oscillators with a chaotic stimulus in panels (a) and (b) and with a periodic stimulus in panels (c) and (d) of Figure \ref{Ross_new}.

\begin{figure}
\begin{center}
\includegraphics[width=7.50cm]{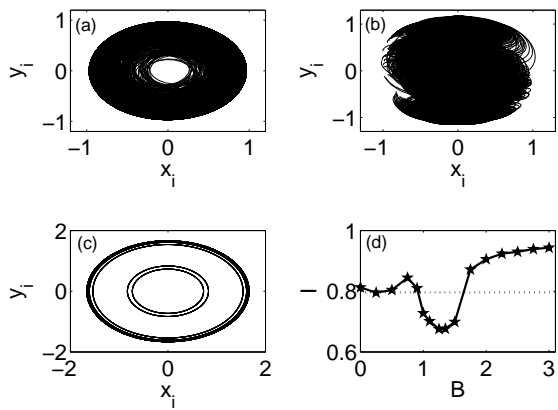}
\caption{Numerical illustration of the occurrence of ERD and ERS in a system of 1000 SL oscillators for different stimulus strengths. Here $A=0.5$, $c=1.5$, $c_e$=2.0, $\omega_e=1.0$, $\gamma=0.05$, $\omega=2.5$ and $B=0.0$ for (a), $B=1.5$ for (b) and $B=3.0$ for (c). (a), (b) and (c) depict the phase portraits in the ($x_i,y_i$) plane. (d) shows the change in the intensity $I$ for different values of $B$, representing ERD and ERS.}
\label{SL_ers_erd}
\end{center}
\end{figure}

The same stimulus can result in ERS and ERD depending upon the strength of the stimulus. Fig. \ref{SL_ers_erd} shows this phenomenon where we choose the initial state of the system to be in a partially synchronized state (since we want to demonstrate both ERS and ERD). Panel (a) represents the initial state before the application of the stimulus. When the stimulus is applied, depending upon the strength of the stimulus $B$ ERD occurs in panel (b) while ERS occurs in panel (c). The change in intensity $I$ for different values of stimulus strength $B$ is shown in panel (d). The reference line corresponds to the initial partial synchronized state and an intensity $I$ above and below the line denotes the occurrence of ERS and ERD, respectively. Thus we find that for a given external stimulus signal, both ERD and ERS can occur depending upon the strength of the stimulus.

\section{Applications}
\label{App}

In this paper we have discussed the occurrence of event-related synchronization and desynchronization in a system of coupled nonlinear oscillators in the presence of an external stimulus. This type of model and the results discussed here can be applied to various real world systems. We point out a few applications below.

Synchronization of neurons at a particular frequency band may facilitate the performance of a particular task while desynchronization at the same frequency band may leave it unattended. The vice verse can also happen. Brain oscillations at different frequency bands are found to be one of the most crucial mechanisms that control higher level information processing, motor functioning and even large scale integration of information across various regions of the brain. 

It is therefore highly important to understand the dynamics of event related neuronal oscillations in relation to specific tasks and/or pathologies. In particular, in pathological conditions like tremors, it is of great importance to understand the occurrence of synchronization or desynchronization of neuronal oscillations at specific frequency bands in order to be able to control the synchronization of neuronal oscillations and hence the pathological states. 

The occurrence of synchronization or desynchronization due to an event is not limited to brain and there are many other systems that exhibit this phenomenon. For instance, an open problem in the field of nano-technology is that the microwave power emitted by a single spin torque nano-oscillator (STNO) is very small, of the order of nano-watts. In order to increase the power one needs to find ways to synchronize a group of coupled nano-oscillators to obtain coherent power. The model discussed in this paper can be used to find ways to avoid desynchronization in such systems and can also explain the occurrence of synchronization and/or desynchronization caused by a stimulated microwave current \cite{Grollier:06}.

We also note here that this model, apart from serving as a qualitative mathematical representative for ERD/ERS phenomena in the brain, can also explain synchronization and/or desynchronization due to external stimulus in other systems; for instance, polariton condensates in semiconductor micro-cavities that interact both among themselves and with the reservoir and quantum coherence of condensates \cite{Szymanska:06,Wouters:08} can also be represented by this model. Studies on these systems will be discussed separately.

Even more generally, the model can account for synchronization/desynchronization due to external stimulus, while the system is already synchronized due to coupling. Possible applications in this direction include the establishment of synchronization or desynchronization due to external stimulus in neuronal networks, Bose--Einstein condensates, Josephson junction arrays and lasers and so on. 

Therefore it will also be of interest to investigate the occurrence of ERD/ERS in systems of diffusively coupled sub-populations. Such a model will facilitate the understanding of the simultaneous occurrence of ERD/ERS at multiple frequency bands in the brain and in systems of coupled STNOs in multiple columns.

\section{Conclusion}
\label{conc}
In this article we have proposed a general model of coupled nonlinear oscillators which can represent interesting physical or biological systems such as a system of coupled spin torque nano-oscillators or neurons, by generalizing and building upon our previous Letter \cite{Sheeba:09}. We subject the system to an external field, which represents an event or an external stimulus. The model results provide an understanding of the occurrence of ERD/ERS due to different types of stimuli. We have found that the occurrence of ERD/ERS is generic to all types of coupled oscillator systems including chaotic systems. Hence we hope that the model results can be applied to quantitative description and prediction of the occurrence of synchronization and desynchronization due to external stimulus in real world systems including Bose--Einstein condensates, Josephson junction arrays and lasers and so on.

The work is supported by the Department of Science and Technology (DST)--Ramanna program, DST--IRHPA research project and a DAE Raja Ramanna program, Government of India. JHS is supported by the DST--FAST TRACK Young Scientist research project.

\end{document}